\documentclass{llncs}

\usepackage{graphicx}
\usepackage{amsmath}
\usepackage{amssymb}
\usepackage{tikz}
\usepackage{algorithm}
\usepackage{algpseudocode}

\usetikzlibrary{arrows,matrix,backgrounds}

\newcommand{\mc}{\mathcal}
\newcommand{\Oh}{{\mc{O}}}
\newcommand{\bigo}{\mathcal{O}}
\newcommand{\bigos}{\mathcal{O}^*}

\newcommand{\NP}{\mbox{$\mathcal{NP}$}}

\newcommand{\DSFULL}{\textsc{Dominating Set}}
\newcommand{\DS}{\textsc{ds}}
\newcommand{\VCFULL}{\textsc{Vertex Cover}}
\newcommand{\VC}{\textsc{vc}}
\newcommand{\MSCFULL}{\textsc{Minimum Set Cover}}

\newcommand{\CEFULL}{\textsc{Cluster Editing}}

\newcommand{\MISFULL}{\textsc{Maximum Independent Set}}

\newcommand{\MCEFULL}{\textsc{Maximal Cliques Enumeration}}

\newcommand{\AL}{\textsc{al}}
\newcommand{\AM}{\textsc{am}}
\newcommand{\COLORSETLIST}{\textsc{color\_set\_list}}
\newcommand{\CSL}{\textsc{csl}}
\newcommand{\IM}{\textsc{im}}
\newcommand{\LIST}{\textsc{list}}
\newcommand{\IDXLIST}{\textsc{idxlist}}
\newcommand{\DEGREE}{\textsc{deg}}
\newcommand{\NDEGREE}{\textsc{ndeg}}
\newcommand{\VCOLOR}{\textsc{vcolor}}
\newcommand{\COLORDEGREE}{\textsc{color\_degree}}
\newcommand{\CD}{\textsc{cd}}
\newcommand{\COLORCARD}{\textsc{color\_card}}
\newcommand{\CC}{\textsc{cc}}

\newcommand{\ALVCOPT}{\textsc{al\_vc\_opt}}
\newcommand{\HYBRIDVCOPT}{\textsc{hybrid\_vc\_opt}}
\newcommand{\HYBRIDVCPARM}{\textsc{hybrid\_vc\_parm}}
\newcommand{\HYBRIDVCFPARM}{\textsc{hybrid\_vcf\_parm}}
\newcommand{\ALDSOPT}{\textsc{al\_ds\_opt}}
\newcommand{\HYBRIDDSOPT}{\textsc{hybrid\_ds\_opt}}
\newcommand{\ALCEPARM}{\textsc{al\_ce\_parm}}
\newcommand{\HYBRIDCEPARM}{\textsc{hybrid\_ce\_parm}}

\renewcommand\footnotemark{}

\begin{document}

\title{A Hybrid Graph Representation for Exact Graph Algorithms
\thanks{A preliminary version of a portion of this paper was presented at
the $4^{th}$ International Frontiers in Algorithmics Workshop, Wuhan, China, 2010.}}

\author
{
    Faisal N. Abu-Khzam\inst{1}
    \and Karim A. Jahed\inst{1}
    \and Amer E. Mouawad\inst{2}
}
\institute
{
    Department of Computer Science and Mathematics\\
    Lebanese American University, Beirut, Lebanon.\\
    \email{\{faisal.abukhzam, karim.jahed\}@lau.edu.lb}
    \and
    David R. Cheriton School of Computer Science\\
    University of Waterloo, Ontario, Canada.\\
    \email{\{aabdomou\}@uwaterloo.ca}
}
\maketitle
\sloppy

%% --------------------------------------------------------------------
%       Abstract
%% --------------------------------------------------------------------

\begin{abstract}
Many exact search algorithms for \NP-hard graph problems adopt the
old Davis-Putman branch-and-reduce paradigm.
The performance of these algorithms often suffers from the increasing number of
graph modifications, such as vertex/edge deletions, that reduce the problem
instance and have to be ``taken back'' frequently during the search process.
We investigate practical implementation-based aspects of exact graph
algorithms by providing a simple hybrid graph representation that trades
space for time to address the said take-back challenge. Experiments on
three well studied problems show
consistent significant improvements over classical methods.
\end{abstract}

\section{Introduction}
Despite their super-polynomial asymptotic running times, exact and parameterized
algorithms for \NP-hard graph problems have recently gained great momentum.
This could be attributed to a number of facts including the hardness of reasonable
polynomial-time approximations as well as the emergence of parameterized
complexity theory.
A great majority of exact graph algorithms adopt the classical search-tree
based recursive backtracking method. The usual goal is to achieve the
least-possible asymptotic worst-case running time.
While this is of some importance from a theoretical standpoint, the practical
significance remains a major challenge.

In general, search-tree based backtracking algorithms for graph
problems employ reduction and
pruning procedures as well as actions associated with branching
decisions that have to be taken, then (frequently) taken back, at every
search-tree node. Such algorithms are often described by simple pseudo-code,
but their implementation could be more sophisticated, mainly due to the
book-keeping needed to be able to undo any graph modifications that result from
reduction procedures.
A main implementation challenge, therefore, is to reduce the additional cost
of {\em undo} operations.

Normally, graph algorithms are implemented with the adjacency list or
the adjacency matrix graph representation. In this paper, we show how a
simple combination of the two representations can be used to facilitate
the {\em undo} of many basic graph operations, especially deletions,
thereby improving the running
time of recursive backtracking (search) algorithms.
Generally, every operation that can be taken back is pushed onto a stack
and later popped out and performed in reverse. We refer to this action by
{\em explicit-undo.} Our implicit-undo objective is achieved via
a highly efficient use of a processor's control stack only.

In addition to effective backtracking, the hybrid representation combines the
advantage of constant time adjacency-queries in adjacency-matrices and the
advantage of efficient neighborhood traversal in adjacency-lists.
This method works well on simple unweighted graphs and may be extended
to weighted graphs. Compared to the use of an adjacency list, the hybrid
representation yields improved running times
for vertex and edge deletion, permanent edge addition, as
well as computing common neighbors.

Despite its simplicity, the introduced method
has a surprising great impact on the implementation of
any recursive backtracking algorithm.
This is shown via experiments conducted on several implementations of known
algorithms, using different techniques that were developed and compared for
three well-known graph problems:
\DSFULL, \VCFULL, and \CEFULL.
The running times of each algorithm is improved by a factor of at least
two on every run, and in some cases the time
was reduced from days to minutes.

The paper is organized as follows. First, some background material is
provided in Section~\ref{sec-background}. The hybrid graph representation is presented and discussed
in Section~\ref{sec-hybrid}. In Section~\ref{sec-operations}, we describe
the effect of the hybrid method on the running time of common graph
modification operations.
Section~\ref{sec-experiments} is devoted to our experimental studies and results, and we
close with some concluding remarks in Section~\ref{sec-conclusion}.

\section{Background}\label{sec-background}
For the sake of completeness, we provide a quick overview
of basic graph representation methods while describing some notation
and terminology adopted in the paper.
An $n$-vertex graph $G = (V, E)$ is usually represented using one of two data structures: adjacency
matrices (\AM) or adjacency lists (\AL).
In this paper, we make use of a degrees' array to keep track of active vertices
and the current cardinalities of their neighborhoods.
When using \AM, neighborhood traversal takes $\Omega(n)$ time.
This is reduced to $\Oh(\Delta(G))$ time, where $\Delta(G)$ is the
maximum degree of $G$, if we use \AL\ instead. On the other hand, checking
if two vertices are adjacent
requires $\Oh(\Delta(G))$ time in \AL\ and $\Oh(1)$ time in \AM.

A vertex cover of a graph $G$ is a set of vertices whose complement induces
an edgeless subgraph.
In the (parameterized) \VCFULL\ problem, or \VC\ for short, we are given a
graph $G = (V, E)$, together with a positive integer $k$,
and we are asked to find a set $C$ of cardinality $k$ such that
$C \subseteq V$ and the subgraph induced by $V \setminus C$ is edgeless. The
current fastest worst-case \VC\ algorithm runs in $\Oh(1.2738^kn^{\bigo(1)})$
time \cite{ChenKanj10}.
An optimization algorithm for \VC\ can be obtained by
obvious modifications to the parameterized algorithm
of \cite{ChenKanj10},
or using the \MISFULL\ algorithm from \cite{FGK06}.

For comparison purposes, four versions were implemented for \VC:
\begin{itemize}
\item \ALVCOPT: an optimization version
using the adjacency-lists representation, based on simple modifications
of the parameterized \VC\ algorithm;

\item \HYBRIDVCOPT: an optimization version using the hybrid graph
representation;

\item \HYBRIDVCPARM: a parameterized version using the hybrid graph
representation but not taking advantage of the {\em folding} technique~\cite{Chen01vertexcover};

\item \HYBRIDVCFPARM: a parameterized version using the hybrid graph
representation and modified for fast edge-contraction operations.
\end{itemize}

In the \DSFULL\ problem, henceforth \DS, we are given an $n$-vertex
graph $G = (V, E)$, and we are asked to find a set $D\subset V$ of smallest
possible cardinality such that every vertex of $G$ is either in $D$ or
adjacent to some vertex in $D$. \DS\ has received great attention, being a
classical \NP-hard graph optimization problem with many logistical applications.

Until 2004, the best algorithm for \DS\ was still the trivial $\Oh^*(2^n)$
enumeration\footnote{Throughout this paper we use the modified big-Oh notation
that suppresses all polynomially bounded factors. For functions $f$ and $g$ we
say $f(n)\in{\Oh^*(g(n))}$ if $f(n)\in{\Oh(g(n)poly(n))}$, where $poly(n)$
is a polynomial.}.
In that same year, two algorithms were independently published breaking the
$\Oh^*(2^n)$ barrier \cite{Fomin04exact,DBLP:journals/jda/Grandoni06}.
The best worst-case algorithm was presented by Grandoni with a running time
in $\Oh^*(1.8019^n)$ \cite{DBLP:journals/jda/Grandoni06}. Using
measure-and-conquer, a bound of $\Oh^*(1.5137^n)$ was obtained on the running
time of Grandoni's algorithm \cite{Fomin05measureand}. This was later
improved to $\Oh^*(1.5063)$ in \cite{conf/stacs/RooijB08} and the current best
worst-case algorithm can be found in \cite{UUCS2008043} where a general
algorithm for counting minimum dominating sets in $\Oh^*(1.5048)$ time is also
presented.

For our experimental work, we implemented two versions of the algorithm
of \cite{Fomin05measureand} where \DS\ is solved by reduction to \MSCFULL:
\begin{itemize}
\item \ALDSOPT: optimization version using the adjacency-lists representation;
\item \HYBRIDDSOPT: optimization version using the hybrid graph representation.
\end{itemize}

Our third example is the \CEFULL\ problem, which takes a graph $G$ and
a positive integer $k$ as input and asks whether deleting or adding a total of
at most $k$ edges yields a transitive graph, i.e., a disjoint union of cliques.
There is a long sequence of fixed-parameter algorithms for this problem,
all are mainly based on dealing with {\em conflict triples}, which are nothing but
induced paths of length two
(see \cite{BBBT09,Bocker2011,BBK11,Cao-Chen,Chen-Meng,Gramm05,Guo09}).
In short, if the graph has an induced path $x-y-z$, then either add the edge $xz$
or delete one of the two other edges. The corresponding branching algorithm
runs in $\bigos(3^k)$ time and can be improved via reduction and pruning methods.
Recent versions of the problem were also shown to yield better running times
by using multiple parameters~\cite{Abu13,KU12}.
In this paper we use the main (generic) algorithm for \CEFULL\ based on
branching on conflict triples, with only simple reduction and pruning. The main
purpose is not to implement a fastest possible algorithm,
but to test the effect of the various operations handled via the
hybrid representation, especially edge addition/deletion.

\section{The Hybrid Graph Representation}\label{sec-hybrid}
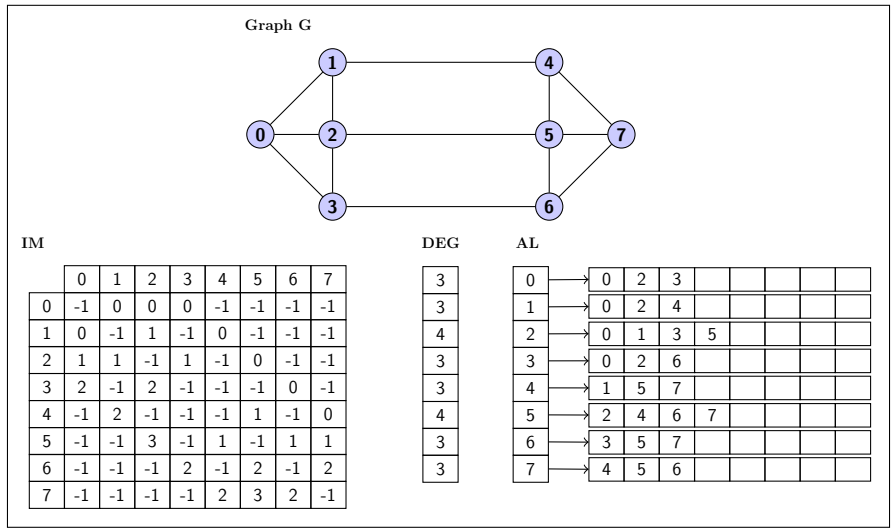
\begin{figure}[htb]
\centering

\begin{tikzpicture}[->,show background rectangle, scale=0.48, every node/.style={anchor=center, scale=0.48},node distance=3cm,main node/.style={circle,fill=blue!20,draw,font=\sffamily\Large\bfseries}]

\node[main node] (0) at (1,0) {0};
\node [main node] (1) at (3,2) {1};
\node [main node] (3) at (3,-2) {3};
\node [main node] (2) at (3,0) {2};
\node [main node] (5) at (9,0) {5};
\node [main node] (7) at (11,0) {7};
\node [main node] (4) at (9,2) {4};
\node [main node] (6) at (9,-2) {6};

\begin{scope}[-]
  \path
  (0) edge (2)
	  edge (1)
      edge (3)
  (1) edge (4)
      edge (2)
  (2) edge (5)
      edge (3)
  (3) edge (6)
  (7) edge (4)
      edge (5)
      edge(6)
  (5) edge (4)
      edge (6);

\matrix [matrix of nodes,
    row sep=-\pgflinewidth,
    column sep=-\pgflinewidth,
    style={font=\sffamily\Large, nodes={rectangle,draw,minimum width=3em,minimum height=2.3em}}] at (-1,-7){
&   & 0 & 1 & 2 & 3 & 4 & 5 & 6 & 7\\
& 0 & -1 & 0 & 0 & 0 & -1 & -1 & -1 & -1\\
& 1 & 0 & -1 & 1 & -1 & 0 & -1 & -1 & -1\\
& 2 & 1 & 1 & -1 & 1 & -1 & 0 & -1 & -1\\
& 3 & 2 & -1 & 2 & -1 & -1 & -1 & 0 & -1\\
& 4 & -1 & 2 & -1 & -1 & -1 & 1 & -1 & 0\\
& 5 & -1 & -1 & 3 & -1 & 1 & -1 & 1 & 1\\
& 6 & -1 & -1 & -1 & 2 & -1 & 2 & -1 & 2\\
&7 & -1 & -1 & -1 & -1 & 2 & 3 & 2 & -1\\
};

\matrix [matrix of nodes,
    row sep=-\pgflinewidth,
    column sep=-\pgflinewidth,
    style={font=\sffamily\Large, nodes={rectangle,draw,minimum width=3em,minimum height=2.3em}}] at (6,-6.65){
& 3\\
& 3\\
& 4\\
& 3\\
& 3\\
& 4\\
& 3\\
& 3\\
};

\matrix [matrix of nodes,
    row sep=-\pgflinewidth,
    column sep=-\pgflinewidth,
    style={font=\sffamily\Large, nodes={rectangle,draw,minimum width=3em,minimum height=2.3em}}] at (8.5,-6.65){
& \node[] (al_point_1) {0};\\
& \node[] (al_point_2) {1};\\
& \node[] (al_point_3) {2};\\
& \node[] (al_point_4) {3};\\
& \node[] (al_point_5) {4};\\
& \node[] (al_point_6) {5};\\
& \node[] (al_point_7) {6};\\
& \node[] (al_point_8) {7};\\
};

\matrix [matrix of nodes,
    row sep=1,
    column sep=-\pgflinewidth,
    style={nodes in empty cells, font=\sffamily\Large, nodes={rectangle,draw,minimum width=3em,minimum height=2em}}] at (14,-6.65){
\node[] (al_row_1) {0}; & 2 & 3 & & & & &\\
\node[] (al_row_2) {0}; & 2 & 4 & & & & &\\
\node[] (al_row_3) {0}; & 1 & 3 & 5 & & & &\\
\node[] (al_row_4) {0}; & 2 & 6 & & & & &\\
\node[] (al_row_5) {1}; & 5 & 7 & & & & &\\
\node[] (al_row_6) {2}; & 4 & 6 & 7 & & & &\\
\node[] (al_row_7) {3}; & 5 & 7 & & & & &\\
\node[] (al_row_8) {4}; & 5 & 6 & & & & &\\
};
\end{scope}

\path
  (al_point_1) edge (al_row_1)
  (al_point_2) edge (al_row_2)
  (al_point_3) edge (al_row_3)
  (al_point_4) edge (al_row_4)
  (al_point_5) edge (al_row_5)
  (al_point_6) edge (al_row_6)
  (al_point_7) edge (al_row_7)
  (al_point_8) edge (al_row_8);

\node[] at (1.5,3) {\large \bf Graph G};
\node at (-5.3,-3.0) {\large \bf IM};
\node at (6,-3.0) {\large \bf DEG};
\node at (8.4,-3.0) {\large \bf  AL};
\end{tikzpicture}

\caption{Example graph G and the corresponding hybrid representation.}
\label{hybridfig}
\end{figure}

The hybrid graph representation is best described using a simple
example such as the one given in Figure \ref{hybridfig}.
The adjacency list of a vertex $v$ is stored in an array denoted by
$\AL[v]$. Accordingly, $\AL[v][i]$ holds the index of the $i^{th}$ vertex in the list
of neighbors of $v$. The adjacency matrix, denoted
henceforth by \IM, is used as an index table for the adjacency list as follows:
the entry $\IM[u][v]$ is equal to the index of $u$ in $\AL[v]$. $\IM[u][v]$ is
set to $-1$ when the two vertices are not adjacent.

Considering the $8$-vertex graph $G = (V,E)$, the initial contents of \AL\ and \IM\
are shown in Figure \ref{hybridfig}.
Note that \AL\ is implemented using a two dimensional array for fast (direct)
access via the indexing provided by \IM. In general, we allocate
enough memory to fit the neighbors of each vertex only.
In addition to \IM\ and \AL, we introduce three linear arrays:
the degree vector (\DEGREE),
the vertex list (\LIST), and the vertex index list (\IDXLIST).

The degree vector holds the current neighborhood cardinality of each vertex, \LIST\
contains the list of currently active (not deleted) vertices which we use instead
of the degree vector for more efficient complete graph traversals, and \IDXLIST\
stores the index of each vertex in \LIST.
In other words, $\LIST[i]$ is the $i^{th}$ vertex in the list of active vertices and
$\IDXLIST[u]$ is the index of vertex $u$ in \LIST.
All data structures except for the \DEGREE\ vector are global and their memory is
allocated at startup only.
The \DEGREE\ vector is local to every search-tree node (i.e., every recursive
call, since a copy of the vector is passed as parameter).

In the next section, we show how the above structures are dynamically updated during
a search algorithm, while performing various graph modification operations. We note that
some operations, like edge contraction for example, require
additional bookkeeping that we briefly describe later. However, most common
operations can be performed using the five data structures described
above, which when combined together form the (generic) hybrid graph representation.

\section{Efficient Search Operations}\label{sec-operations}
The hybrid data structure was designed with an eye on efficient graph modification
operations during backtracking. In addition, some basic operations like adjacency
query and neighborhood traversal are also performed efficiently. For example, checking
if two vertices $u$ and $v$ are adjacent takes $\Oh(1)$ time:
just check whether $-1 < \IM[u][v] < \DEGREE[v]$.
Neighborhood search, on the other hand, runs in $\Oh(d)$ time, where $d$ is the
current vertex degree.

Graph traversal is another frequent operation, often used to select a vertex or an
edge having a certain property during branching algorithms.
A complete graph traversal runs in $\Oh(n_c)$ time as opposed to
$\Oh(n)$, where $n_c$ is the number of currently active vertices in the graph.
This is possible
because of our use of the \LIST\ and \IDXLIST\ vectors. In the rest of this section,
we discuss how the hybrid method is used for graph modification operations.

\subsection{Edge Deletion}
The simplest and most frequent operation performed during the search is probably edge
deletion. Maintaining the hybrid data structure in this case is straightforward. For
deleting an edge ($u$, $v$), the degrees of $u$ and $v$ are decremented by one and the
adjacency lists of the two vertices are adjusted respectively by placing $u$
at the last position of $\AL[v]$ and $v$ at the last position of $\AL[u]$ (Figure~\ref{proc-delete-edge}). Each of
these two operations consists of a single swap with the last element of the
respective list, together with an adjustment of the positions in \IM.

\begin{figure}[!t]
\centering
\hrulefill
\begin{algorithmic}[1]
\Procedure{DeleteEdge}{$u$,$v$}
  \State $i \gets \IM[v][u]$;
  \State $j \gets \DEGREE[u] - 1$;
  \State $x \gets \AL[u][j]$;
  \State $\AL[u][i] \gets x$;
  \State $\AL[u][j] \gets v$;
  \State $\IM[x][u] \gets i$;
  \State $\IM[v][u] \gets j$;
  \State $\DEGREE[u] \gets \DEGREE[u] - 1$;
  \State
  \State $i \gets \IM[u][v]$;
  \State $j \gets \DEGREE[v] - 1$;
  \State $x \gets \AL[v][j]$;
  \State $\AL[v][i] \gets x$;
  \State $\AL[v][j] \gets u$;
  \State $\IM[x][v] \gets i$;
  \State $\IM[u][v] \gets j$;
  \State $\DEGREE[v] \gets \DEGREE[v] - 1$;
\EndProcedure
\end{algorithmic}
\hrulefill
\caption{The {\sc DeleteEdge} procedure.}
\label{proc-delete-edge}
\end{figure}

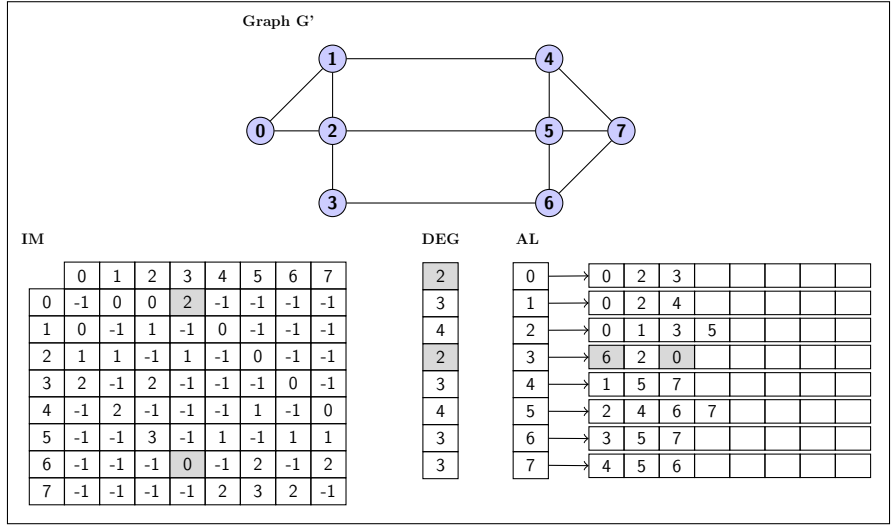
\begin{figure}[htb]
\centering

\begin{tikzpicture}[->,show background rectangle, scale=0.48, every node/.style={anchor=center, scale=0.48},node distance=3cm,main node/.style={circle,fill=blue!20,draw,font=\sffamily\Large\bfseries}]

\node[main node] (0) at (1,0) {0};
\node [main node] (1) at (3,2) {1};
\node [main node] (3) at (3,-2) {3};
\node [main node] (2) at (3,0) {2};
\node [main node] (5) at (9,0) {5};
\node [main node] (7) at (11,0) {7};
\node [main node] (4) at (9,2) {4};
\node [main node] (6) at (9,-2) {6};

\begin{scope}[-]
  \path
  (0) edge (2)
	  edge (1)
  (1) edge (4)
      edge (2)
  (2) edge (5)
      edge (3)
  (3) edge (6)
  (7) edge (4)
      edge (5)
      edge(6)
  (5) edge (4)
      edge (6);

\matrix [matrix of nodes,
    row sep=-\pgflinewidth,
    column sep=-\pgflinewidth,
    style={font=\sffamily\Large, nodes={rectangle,draw,minimum width=3em,minimum height=2.3em}}] at (-1,-7){
&   & 0 & 1 & 2 & 3 & 4 & 5 & 6 & 7\\
& 0 & -1 & 0 & 0 & \node[style={fill=gray!30}] {2}; & -1 & -1 & -1 & -1\\
& 1 & 0 & -1 & 1 & -1 & 0 & -1 & -1 & -1\\
& 2 & 1 & 1 & -1 & 1 & -1 & 0 & -1 & -1\\
& 3 & 2 & -1 & 2 & -1 & -1 & -1 & 0 & -1\\
& 4 & -1 & 2 & -1 & -1 & -1 & 1 & -1 & 0\\
& 5 & -1 & -1 & 3 & -1 & 1 & -1 & 1 & 1\\
& 6 & -1 & -1 & -1 & \node[style={fill=gray!30}] {0}; & -1 & 2 & -1 & 2\\
&7 & -1 & -1 & -1 & -1 & 2 & 3 & 2 & -1\\
};

\matrix [matrix of nodes,
    row sep=-\pgflinewidth,
    column sep=-\pgflinewidth,
    style={font=\sffamily\Large, nodes={rectangle,draw,minimum width=3em,minimum height=2.3em}}] at (6,-6.65){
& \node[style={fill=gray!30}] {2};\\
& 3\\
& 4\\
& \node[style={fill=gray!30}] {2};\\
& 3\\
& 4\\
& 3\\
& 3\\
};

\matrix [matrix of nodes,
    row sep=-\pgflinewidth,
    column sep=-\pgflinewidth,
    style={font=\sffamily\Large, nodes={rectangle,draw,minimum width=3em,minimum height=2.3em}}] at (8.5,-6.65){
& \node[] (al_point_1) {0};\\
& \node[] (al_point_2) {1};\\
& \node[] (al_point_3) {2};\\
& \node[] (al_point_4) {3};\\
& \node[] (al_point_5) {4};\\
& \node[] (al_point_6) {5};\\
& \node[] (al_point_7) {6};\\
& \node[] (al_point_8) {7};\\
};

\matrix [matrix of nodes,
    row sep=1,
    column sep=-\pgflinewidth,
    style={nodes in empty cells, font=\sffamily\Large, nodes={rectangle,draw,minimum width=3em,minimum height=2em}}] at (14,-6.65){
\node[] (al_row_1) {0}; & 2 & 3 & & & & &\\
\node[] (al_row_2) {0}; & 2 & 4 & & & & &\\
\node[] (al_row_3) {0}; & 1 & 3 & 5 & & & &\\
\node[style={fill=gray!30}] (al_row_4) {6}; & 2 & \node[style={fill=gray!30}] {0}; & & & & &\\
\node[] (al_row_5) {1}; & 5 & 7 & & & & &\\
\node[] (al_row_6) {2}; & 4 & 6 & 7 & & & &\\
\node[] (al_row_7) {3}; & 5 & 7 & & & & &\\
\node[] (al_row_8) {4}; & 5 & 6 & & & & &\\
};
\end{scope}

\path
  (al_point_1) edge (al_row_1)
  (al_point_2) edge (al_row_2)
  (al_point_3) edge (al_row_3)
  (al_point_4) edge (al_row_4)
  (al_point_5) edge (al_row_5)
  (al_point_6) edge (al_row_6)
  (al_point_7) edge (al_row_7)
  (al_point_8) edge (al_row_8);

\node[] at (1.5,3) {\large \bf Graph G'};
\node at (-5.3,-3.0) {\large \bf IM};
\node at (6,-3.0) {\large \bf DEG};
\node at (8.4,-3.0) {\large \bf  AL};
\end{tikzpicture}
\caption{Hybrid representation after deleting edge ($v_0$, $v_3$).}
\label{hybridfig2}
\end{figure}

Going back to our illustrative graph $G$, after deleting edge ($v_0$, $v_3$), the
modified $\AL$, $\IM$, and $\DEGREE$ vectors will be as shown in
Figure \ref{hybridfig2} (changes are shown in gray).
No changes to \LIST\ and \IDXLIST\ are required for edge deletion since no
vertices are removed from the graph.

Notice that this operation runs in $\Oh(1)$ time as all the information
required for switching positions in \AL\ can be found in \IM. Vertex $v_3$ is
no longer a neighbor of vertex $v_0$
because $\IM[0][3] = 2$ which is not less than $\DEGREE[0] = 2$.

\subsection{Undo Edge Deletion}
Now assume we want to undo the previous operation. This can be accomplished
by simply setting $\DEGREE[0]$ and $\DEGREE[3]$ back to 3.
There is no need to change \IM\ and \AL\ back to the previous state
since no specific order is required for storing vertex neighbors.
Moreover, since every search-tree node maintains its own copy of
the \DEGREE\ vector, there are actually no actions whatsoever that need to be
taken for undoing edge deletion. Thus we have an {\em implicit-undo} for
edge deletion operations. This implicit
undo runs in $\Oh(1)$ time.

\subsection{Vertex Deletion}
In a typical computation using adjacency lists, deleting a vertex $v$ of current
degree $d$ requires traversing the list of neighbors of $v$ and deleting
$v$ from the list of each of its neighbors. This operation runs in $\bigo(\Delta(G)^2)$ time.

Using our hybrid computation,
deleting a vertex $v$ runs in $\bigo(d)$ time where $d$ is the (current) degree of $v$.
This is simply performed by running the
edge deletion operation for every active neighbor of $v$ (Figure~\ref{proc-delete-vertex}). In addition,
we remove $v$ from the list of active vertices by swapping it with the last active
vertex in \LIST\ and decrementing the number of active vertices by one.
The \IDXLIST\ plays the same role as \IM\ for deleting a vertex from \LIST.

To illustrate the purpose of the \LIST\ vector, suppose we need to pick a vertex
with a particular property, such as one of maximum degree
(which is often needed for heuristic priorities).

Consider the two operations of copying
the \DEGREE\ vector and searching for the vertex of highest degree. Doing so would
consume $\Omega(n)$ time if the \DEGREE\ vector is used alone. This is reduced
to $\bigo(n_c)$, where $n_c$ is the number of currently active vertices.
To see why, note that iterating
from $i = 0$ to $n_c$ only, $\DEGREE[\LIST[i]]$ returns the degree of the
vertex at position $i$ in \LIST.
Knowing that a great majority of search tree nodes are near the bottom of a search
tree, where the graph order is almost constant, this simple strategy alone makes
a substantial difference.

\begin{figure}[!t]
\centering
\hrulefill
\begin{algorithmic}[1]
\Procedure{DeleteVertex}{$u$,$n_c$}
  \State $d \gets \DEGREE[v]$;
  \State $last \gets \LIST[n_c - 1]$;
  \State $i \gets \IDXLIST[v]$;
  \State $\LIST[i] \gets last$;
  \State $\LIST[n_c - 1] \gets v$;
  \State $\IDXLIST[last] \gets i$;
  \State $\IDXLIST[v] \gets n_c - 1$;
  \For{$j \gets d-1, 0$}
    \State $u \gets \AL[v][i]$;
    \State \Call{DeleteEdge}{$u$,$v$};
  \EndFor
\EndProcedure
\end{algorithmic}
\hrulefill
\caption{The {\sc DeleteVertex} procedure.}
\label{proc-delete-vertex}
\end{figure}

\subsection{Undo Vertex Deletion}
As in the case of edge-deletion, the undo of vertex deletion is implicit
(requires no action)
when separate degree vectors are maintained at every search node. This
is due to the fact that edge deletion
can be considered as an independent operation.
The only required operation is to increment the number of active vertices by one
(if this value is stored globally) so that the \LIST\ and \IDXLIST\ vectors
reflect the re-insertion of a vertex.
Explicitly undoing a vertex deletion operation runs in $\Oh(d)$ time
where $d$ is the degree of the deleted vertex (it mainly consists of
incrementing the degree of each neighbor by one.)

\subsection{Edge Contraction}
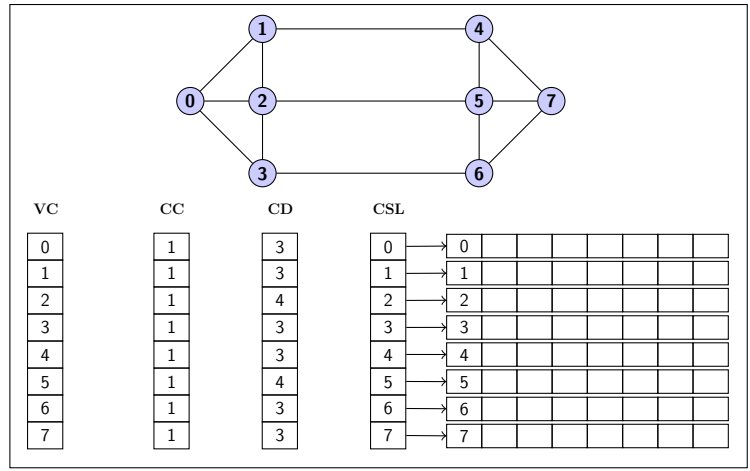
\begin{figure}[htb]
\centering

\begin{tikzpicture}[->,show background rectangle, scale=0.48, every node/.style={anchor=center, scale=0.48},node distance=3cm,main node/.style={circle,fill=blue!20,draw,font=\sffamily\Large\bfseries}]

\node[main node] (0) at (3,0) {0};
\node [main node] (1) at (5,2) {1};
\node [main node] (3) at (5,-2) {3};
\node [main node] (2) at (5,0) {2};
\node [main node] (5) at (11,0) {5};
\node [main node] (7) at (13,0) {7};
\node [main node] (4) at (11,2) {4};
\node [main node] (6) at (11,-2) {6};

\begin{scope}[-]
  \path
  (0) edge (2)
	  edge (1)
      edge (3)
  (1) edge (4)
      edge (2)
  (2) edge (5)
      edge (3)
  (3) edge (6)
  (7) edge (4)
      edge (5)
      edge(6)
  (5) edge (4)
      edge (6);

\matrix [matrix of nodes,
    row sep=-\pgflinewidth,
    column sep=-\pgflinewidth,
    style={font=\sffamily\Large, nodes={rectangle,draw,minimum width=3em,minimum height=2.3em}}] at (-1,-6.65){
& 0\\
& 1\\
& 2\\
& 3\\
& 4\\
& 5\\
& 6\\
& 7\\
};

\matrix [matrix of nodes,
    row sep=-\pgflinewidth,
    column sep=-\pgflinewidth,
    style={font=\sffamily\Large, nodes={rectangle,draw,minimum width=3em,minimum height=2.3em}}] at (2.5,-6.65){
& 1\\
& 1\\
& 1\\
& 1\\
& 1\\
& 1\\
& 1\\
& 1\\
};

\matrix [matrix of nodes,
    row sep=-\pgflinewidth,
    column sep=-\pgflinewidth,
    style={font=\sffamily\Large, nodes={rectangle,draw,minimum width=3em,minimum height=2.3em}}] at (5.5,-6.65){
& 3\\
& 3\\
& 4\\
& 3\\
& 3\\
& 4\\
& 3\\
& 3\\
};

\matrix [matrix of nodes,
    row sep=-\pgflinewidth,
    column sep=-\pgflinewidth,
    style={font=\sffamily\Large, nodes={rectangle,draw,minimum width=3em,minimum height=2.3em}}] at (8.5,-6.65){
& \node[] (al_point_1) {0};\\
& \node[] (al_point_2) {1};\\
& \node[] (al_point_3) {2};\\
& \node[] (al_point_4) {3};\\
& \node[] (al_point_5) {4};\\
& \node[] (al_point_6) {5};\\
& \node[] (al_point_7) {6};\\
& \node[] (al_point_8) {7};\\
};

\matrix [matrix of nodes,
    row sep=1,
    column sep=-\pgflinewidth,
    style={nodes in empty cells, font=\sffamily\Large, nodes={rectangle,draw,minimum width=3em,minimum height=2em}}] at (14,-6.65){
\node[] (al_row_1) {0}; & & & & & & &\\
\node[] (al_row_2) {1}; & & & & & & &\\
\node[] (al_row_3) {2}; & & & & & & &\\
\node[] (al_row_4) {3}; & & & & & & &\\
\node[] (al_row_5) {4}; & & & & & & &\\
\node[] (al_row_6) {5}; & & & & & & &\\
\node[] (al_row_7) {6}; & & & & & & &\\
\node[] (al_row_8) {7}; & & & & & & &\\
};
\end{scope}

\path
  (al_point_1) edge (al_row_1)
  (al_point_2) edge (al_row_2)
  (al_point_3) edge (al_row_3)
  (al_point_4) edge (al_row_4)
  (al_point_5) edge (al_row_5)
  (al_point_6) edge (al_row_6)
  (al_point_7) edge (al_row_7)
  (al_point_8) edge (al_row_8);

\node at (-1,-3) {\large \bf VC};
\node at (2.5,-3) {\large \bf CC};
\node at (5.5,-3) {\large \bf  CD};
\node at (8.5,-3) {\large \bf  CSL};
\end{tikzpicture}

\caption{Coloring data structures (previous data structures not shown but required).}
\label{coloring}
\end{figure}

The next operation we consider is edge contraction.
Contracting edge $(u,v)$ replaces vertices $u$ and $v$ by a new vertex
whose neighborhood is $N(u) \cup N(v) \setminus \{u, v\}$. To
achieve our implicit-undo objective, we implement this
operation using a coloring technique that
requires additional bookkeeping. Simply, vertices with the same color
are treated as one single vertex obtained by contracting edges
between them.
Initially, every vertex $v_i$ is assigned
color $c_i$, and every color class $c_i$ has cardinality one and
degree $d(c_i) = d(v_i)$. In addition to previously discussed data structures,
we use the following (see Figure \ref{coloring}):

\begin{itemize}
\item \VCOLOR\ vector: holds the current color of every vertex.
\item \COLORCARD\ (\CC) vector: holds the current cardinality of every color set.
\item \COLORDEGREE\ (\CD) vector: holds the current degree of every color set.
\item \COLORSETLIST\ (\CSL): holds the list of vertices belonging to every color set.
\end{itemize}

\noindent Given that all operations now involve color sets,
the \LIST\ and \IDXLIST\ do not hold vertex
information anymore, but they maintain the list of active (not deleted)
colors instead.
If no edge contraction operations are performed, color sets would be
identified by their corresponding vertices (or vice-versa).

\begin{figure}[htb]
\centering

\begin{tikzpicture}[->,show background rectangle, scale=0.48, every node/.style={ anchor=center, scale=0.48},node distance=3cm,main node/.style={circle,fill=blue!20,draw,font=\sffamily\Large\bfseries}]

\node[main node] (0) at (3,0) {0};
\node [main node] (1) at (5,2) {1};
\node [main node, style={text width=1.3em}] (3) at (5,-2) {};
\node [main node] (2) at (5,0) {2};
\node [main node] (5) at (11,0) {5};
\node [main node] (7) at (13,0) {7};
\node [main node] (4) at (11,2) {4};
\node [main node, style={text width=1.3em}] (6) at (11,-2){};

\draw  (8,-2) ellipse (4 and 0.7);

\begin{scope}[-]
  \path
  (0) edge (2)
	  edge (1)
      edge (3)
  (1) edge (4)
      edge (2)
  (2) edge (5)
      edge (3)
  (7) edge (4)
      edge (5)
      edge(6)
  (5) edge (4)
      edge (6);

\matrix [matrix of nodes,
    row sep=-\pgflinewidth,
    column sep=-\pgflinewidth,
    style={font=\sffamily\Large, nodes={rectangle,draw,minimum width=3em,minimum height=2.3em}}] at (-1,-6.65){
& 0\\
& 1\\
& 2\\
& 3\\
& 4\\
& 5\\
& \node[style={fill=gray!30}] {3};\\
& 7\\
};

\matrix [matrix of nodes,
    row sep=-\pgflinewidth,
    column sep=-\pgflinewidth,
    style={font=\sffamily\Large, nodes={rectangle,draw,minimum width=3em,minimum height=2.3em}}] at (2.5,-6.65){
& 1\\
& 1\\
& 1\\
& \node[style={fill=gray!30}] {2};\\
& 1\\
& 1\\
& \node[style={fill=gray!30}] {0};\\
& 1\\
};

\matrix [matrix of nodes,
    row sep=-\pgflinewidth,
    column sep=-\pgflinewidth,
    style={font=\sffamily\Large, nodes={rectangle,draw,minimum width=3em,minimum height=2.3em}}] at (5.5,-6.65){
& 3\\
& 3\\
& 4\\
& \node[style={fill=gray!30}] {4};\\
& 3\\
& 4\\
& \node[style={fill=gray!30}] {0};\\
& 3\\
};

\matrix [matrix of nodes,
    row sep=-\pgflinewidth,
    column sep=-\pgflinewidth,
    style={font=\sffamily\Large, nodes={rectangle,draw,minimum width=3em,minimum height=2.3em}}] at (8.5,-6.65){
& \node[] (al_point_1) {0};\\
& \node[] (al_point_2) {1};\\
& \node[] (al_point_3) {2};\\
& \node[] (al_point_4) {3};\\
& \node[] (al_point_5) {4};\\
& \node[] (al_point_6) {5};\\
& \node[] (al_point_7) {6};\\
& \node[] (al_point_8) {7};\\
};

\matrix [matrix of nodes,
    row sep=1,
    column sep=-\pgflinewidth,
    style={nodes in empty cells, font=\sffamily\Large, nodes={rectangle,draw,minimum width=3em,minimum height=2em}}] at (14,-6.65){
\node[] (al_row_1) {0}; & & & & & & &\\
\node[] (al_row_2) {1}; & & & & & & &\\
\node[] (al_row_3) {2}; & & & & & & &\\
\node[] (al_row_4) {3}; & \node[style={fill=gray!30}] {6}; & & & & & &\\
\node[] (al_row_5) {4}; & & & & & & &\\
\node[] (al_row_6) {5}; & & & & & & &\\
\node[] (al_row_7) {6}; & & & & & & &\\
\node[] (al_row_8) {7}; & & & & & & &\\
};
\end{scope}

\path
  (al_point_1) edge (al_row_1)
  (al_point_2) edge (al_row_2)
  (al_point_3) edge (al_row_3)
  (al_point_4) edge (al_row_4)
  (al_point_5) edge (al_row_5)
  (al_point_6) edge (al_row_6)
  (al_point_7) edge (al_row_7)
  (al_point_8) edge (al_row_8);

\node at (8,-2) {\Large \bf 3};
\node at (-1,-3) {\large \bf VC};
\node at (2.5,-3) {\large \bf CC};
\node at (5.5,-3) {\large \bf  CD};
\node at (8.5,-3) {\large \bf  CSL};
\end{tikzpicture}
\caption{Coloring data structures after contracting edge $(v_3, v_6)$ (modifications to previous data structures not shown here but required).}
\label{coloring2}
\end{figure}
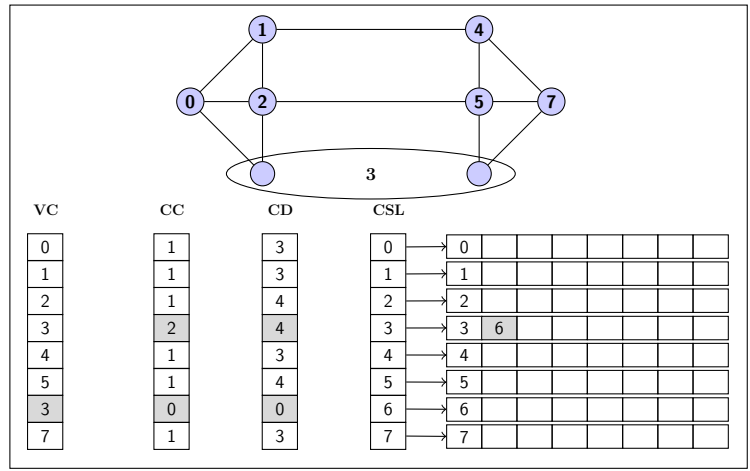

When edge contraction is possible, the initial state of the graph $G$
consists of all data structures previously discussed.
\AL, \IM, \CSL, \LIST\ and \IDXLIST\ would be globally stored
(in RAM), while \DEGREE, \CD, \CC\ and \VCOLOR\ would be copied at every
search-tree node.
To contract an edge, say $(v_3, v_6)$, we actually assign both vertices
the same color. Assuming
we assign the two vertices color $c_3$, the required modifications are shown in
Figure \ref{coloring2} (changes shown in gray).

Since we are dealing with simple graphs, any edge between two vertices belonging
to the same color set is deleted and no more than one edge is allowed between
a color set and another. Clearly, such an operation can be implicitly taken back
as well, but is also more time and space consuming than simple vertex deletion.
This technique makes it possible to implement the vertex folding operation,
introduced in \cite{Chen01vertexcover}, for the parameterized \VC\ algorithm.

\subsection{The Vertex Cover Folding Operation}
Let $(G,k)$ be an instance of the \VCFULL\ problem and let $u\in V(G)$ be
a degree-two vertex with neighbors $v$ and $w$. If $v$ and $w$ are adjacent, then
there is a minimum vertex cover that contains $v$ and $w$ (and not $u$). So it is
safe to delete $u$, add $v$ and $w$ to the potential solution and decrement $k$ by two.
In the case where $v$ and $w$ are non-adjacent,
an equivalent \VCFULL\ instance is obtained by contracting edges $uv$ and
$uw$ and decrementing $k$ by one. This latter operation is known as degree-two
vertex folding \cite{Chen01vertexcover}.

As we shall see, applying the coloring technique
to implement vertex folding considerably improves the runtime on certain recalcitrant
instances, but slows down the computation on
graphs where folding rarely occurs. In such cases, the overhead of maintaining
color-sets is a drawback. Note that folding alone made it possible to obtain
a worst-case running time of $\bigos(1.285^k)$ in \cite{Chen01vertexcover}. Yet, our results
show that excluding folding from the same algorithm is faster on a large number of
instances, except for the case where is graph is regular (or nearly regular) where
other reduction methods do no apply.

\subsection{Permanent Edge Addition}
An edge addition performed at a search-tree node is considered permanent
if no descendant search-tree node later modifies the operation
(by edge or vertex deletion).
In some problems, such as \CEFULL, searching for a solution
involves edge deletions as well as permanent edge additions.

Using the classical adjacency list structure, it would be impossible
to incorporate edge addition without extra bookkeeping to record changes
to the \AL.
To incorporate the permanent edge addition
operation to our hybrid data structure, we make the following modifications:
(i) enough global memory is allocated
to \AL\ (i.e. n-by-n matrix where $n$ is the size of the input graph),
and (ii) an additional degree vector (\NDEGREE) is maintained at every search
node.

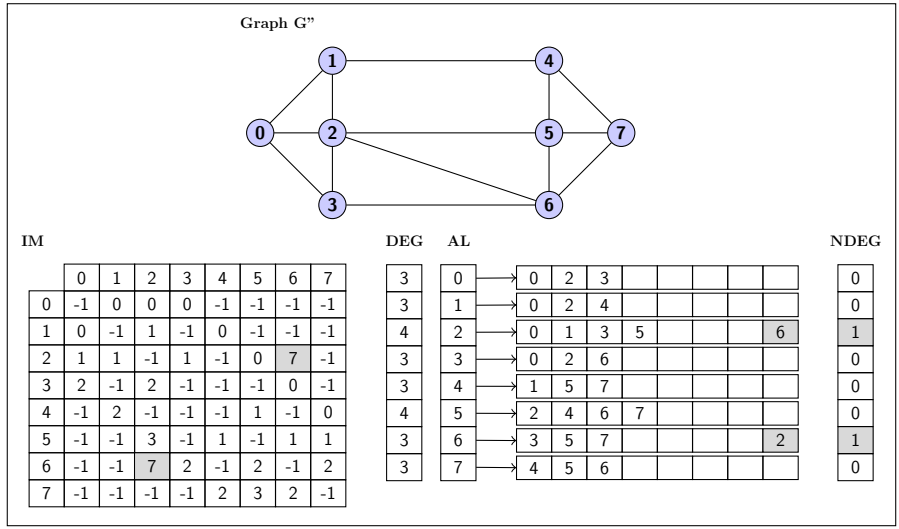
\begin{figure}[htb]
\centering

\begin{tikzpicture}[->,show background rectangle, scale=0.48, every node/.style={anchor=center, scale=0.48},node distance=3cm,main node/.style={circle,fill=blue!20,draw,font=\sffamily\Large\bfseries}]

\node[main node] (0) at (1,0) {0};
\node [main node] (1) at (3,2) {1};
\node [main node] (3) at (3,-2) {3};
\node [main node] (2) at (3,0) {2};
\node [main node] (5) at (9,0) {5};
\node [main node] (7) at (11,0) {7};
\node [main node] (4) at (9,2) {4};
\node [main node] (6) at (9,-2) {6};

\begin{scope}[-]
  \path
  (0) edge (2)
	  edge (1)
      edge (3)
  (1) edge (4)
      edge (2)
  (2) edge (5)
      edge (3)
      edge (6)
  (3) edge (6)
  (7) edge (4)
      edge (5)
      edge(6)
  (5) edge (4)
      edge (6);

\matrix [matrix of nodes,
    row sep=-\pgflinewidth,
    column sep=-\pgflinewidth,
    style={font=\sffamily\Large, nodes={rectangle,draw,minimum width=3em,minimum height=2.3em}}] at (-1,-7){
&   & 0 & 1 & 2 & 3 & 4 & 5 & 6 & 7\\
& 0 & -1 & 0 & 0 & 0 & -1 & -1 & -1 & -1\\
& 1 & 0 & -1 & 1 & -1 & 0 & -1 & -1 & -1\\
& 2 & 1 & 1 & -1 & 1 & -1 & 0 & \node[style={fill=gray!30}] {7}; & -1\\
& 3 & 2 & -1 & 2 & -1 & -1 & -1 & 0 & -1\\
& 4 & -1 & 2 & -1 & -1 & -1 & 1 & -1 & 0\\
& 5 & -1 & -1 & 3 & -1 & 1 & -1 & 1 & 1\\
& 6 & -1 & -1 & \node[style={fill=gray!30}] {7}; & 2 & -1 & 2 & -1 & 2\\
&7 & -1 & -1 & -1 & -1 & 2 & 3 & 2 & -1\\
};

\matrix [matrix of nodes,
    row sep=-\pgflinewidth,
    column sep=-\pgflinewidth,
    style={font=\sffamily\Large, nodes={rectangle,draw,minimum width=3em,minimum height=2.3em}}] at (5,-6.65){
& 3\\
& 3\\
& 4\\
& 3\\
& 3\\
& 4\\
& 3\\
& 3\\
};

\matrix [matrix of nodes,
    row sep=-\pgflinewidth,
    column sep=-\pgflinewidth,
    style={font=\sffamily\Large, nodes={rectangle,draw,minimum width=3em,minimum height=2.3em}}] at (6.5,-6.65){
& \node[] (al_point_1) {0};\\
& \node[] (al_point_2) {1};\\
& \node[] (al_point_3) {2};\\
& \node[] (al_point_4) {3};\\
& \node[] (al_point_5) {4};\\
& \node[] (al_point_6) {5};\\
& \node[] (al_point_7) {6};\\
& \node[] (al_point_8) {7};\\
};

\matrix [matrix of nodes,
    row sep=1,
    column sep=-\pgflinewidth,
    style={nodes in empty cells, font=\sffamily\Large, nodes={rectangle,draw,minimum width=3em,minimum height=2em}}] at (12,-6.65){
\node[] (al_row_1) {0}; & 2 & 3 & & & & &\\
\node[] (al_row_2) {0}; & 2 & 4 & & & & &\\
\node[] (al_row_3) {0}; & 1 & 3 & 5 & & & & \node[style={fill=gray!30}] {6};\\
\node[] (al_row_4) {0}; & 2 & 6 & & & & &\\
\node[] (al_row_5) {1}; & 5 & 7 & & & & &\\
\node[] (al_row_6) {2}; & 4 & 6 & 7 & & & &\\
\node[] (al_row_7) {3}; & 5 & 7 & & & & & \node[style={fill=gray!30}] {2};\\
\node[] (al_row_8) {4}; & 5 & 6 & & & & &\\
};

\matrix [matrix of nodes,
    row sep=-\pgflinewidth,
    column sep=-\pgflinewidth,
    style={font=\sffamily\Large, nodes={rectangle,draw,minimum width=3em,minimum height=2.3em}}] at (17.5,-6.65){
& 0\\
& 0\\
& \node[style={fill=gray!30}] {1};\\
& 0\\
& 0\\
& 0\\
& \node[style={fill=gray!30}] {1};\\
& 0\\
};

\end{scope}

\path
  (al_point_1) edge (al_row_1)
  (al_point_2) edge (al_row_2)
  (al_point_3) edge (al_row_3)
  (al_point_4) edge (al_row_4)
  (al_point_5) edge (al_row_5)
  (al_point_6) edge (al_row_6)
  (al_point_7) edge (al_row_7)
  (al_point_8) edge (al_row_8);

\node[] at (1.5,3) {\large \bf Graph G''};
\node at (-5.3,-3.0) {\large \bf IM};
\node at (5,-3.0) {\large \bf DEG};
\node at (6.5,-3.0) {\large \bf  AL};
\node at (17.5,-3.0) {\large \bf  NDEG};
\end{tikzpicture}

\caption{Extended hybrid representation after adding edge ($v_2$, $v_6$).}
\label{addfig}
\end{figure}

The additional memory requirement guarantees storage space
and the \NDEGREE\ vector serves as a second degree vector.
Figure \ref{addfig} illustrate the required changes to the extended hybrid data
structure after adding edge ($v_2$, $v_6$) to our original graph $G$.

\noindent The adjacency query still takes $\Oh(1)$ time but requires
an additional condition: $u$ and $v$ are
adjacent when  $-1 < \IM[u][v] < \DEGREE[v]$ or $-1 < n - 1 - \IM[u][v] < \NDEGREE[v]$.
Neighborhood traversal becomes slightly more complicated but the time complexity
remains unchanged.
Undoing this operation only requires setting $\NDEGREE[2]$ and $\NDEGREE[6]$ back
to 0 and subsequent addition of edges incident on $v_2$ or $v_6$ could overwrite
older positions.

\subsection{Theoretical Runtime Analysis}
Since we are dealing with exact algorithms for \NP-hard problems, our
target is to achieve faster running times
rather than accommodating large instances, which would not be possible to solve
anyway.
Yet, despite the apparent disadvantage of increasing the ``global'' space
requirement, our algorithms do save the space needed for extra bookkeeping
for explicit undo of graph modifications.
In addition, we achieve constant-time implicit undo-operations which occur every
time our recursive algorithms hit a backtracking state.

Table \ref{tab:timetable} summarizes the advantages of the hybrid graph
representation over the adjacency list and adjacency matrix representations.

\begin{table}[htb]
\caption{\label{tab:timetable}Runtime comparison for different search-operations using \AL, \AM, and the hybrid graph representation. We denote by $n$ the number of vertices in the graph, by $d(v)$ the degree of a vertex $v$, and $d(c)$ denotes the degree of a color-set $c$.}
\begin{center}
\begin{tabular}{| l | c | c | c |} \hline
   Operation & Adjacency Matrix & Adjacency List & Hybrid \\\hline
	Adjacency Query  & $\Oh(1)$ & $\Oh(d(v))$ & $\Oh(1)$\\\hline
	Neighborhood Traversal  & $\Oh(n)$ & $\Oh(d(v))$ & $\Oh(d(v))$\\\hline
    Edge Deletion & $\Oh(1)$ & $\Oh(d(v))$ & $\Oh(1)$\\\hline
	Undo Edge Deletion & $\Oh(1)$ & $\Oh(1)$ & $\Oh(1)$\\\hline
	Vertex Deletion & $\Oh(n)$ & $\Oh(d(v)^2)$ & $\Oh(d(v))$\\\hline
	Undo Vertex Deletion & $\Oh(d(v))$ & $\Oh(d(v))$ & $\Oh(1)$\\\hline
	Edge Contraction & $\Oh(n)$ & $\Oh(d(v))$ & $\Oh(d(c))$\\\hline
	Undo Edge Contraction & $\Oh(d(v))$ & $\Oh(d(v))$ & $\Oh(1)$\\\hline
	Edge Addition & $\Oh(1)$ & $\Oh(1)$ & $\Oh(1)$\\\hline
	Undo Edge Addition & $\Oh(1)$ & $\Oh(d(v))$ & $\Oh(1)$\\\hline
\end{tabular}
\end{center}\vspace{-0.5cm}
\end{table}

\section{Experimental Results}\label{sec-experiments}
Four different versions were implemented for the \VCFULL\ algorithm.
\ALVCOPT\ and \HYBRIDVCOPT\ are two generic search-tree optimization versions using
the adjacency-list and hybrid graph representations respectively.
In Table \ref{tab:vctable1}, the running times for both versions are reported for
a number of DIMACS graphs.

\begin{table}[htb]
\caption{\label{tab:vctable1}\ALVCOPT\ vs. \HYBRIDVCOPT\ (no folding).}
\begin{center}
\begin{tabular}{| c | c | c | c | c | c |} \hline
    Graph & $|V|$ & $|E|$ & $|C|$ & \ALVCOPT\ & \HYBRIDVCOPT\ \\\hline
    %%brock400\_1.clq & 400 & 59723 & 393 & $<$ 1 sec & $<$ 1 sec\\\hline
    %%brock400\_2.clq & 400 & 59786 & 392 & $<$ 1 sec & $<$ 1 sec\\\hline
    %%brock400\_3.clq & 400 & 59681 & 393 & $<$ 1 sec & $<$ 1 sec\\\hline
    %%brock400\_4.clq & 400 & 59765 & 393 & $<$ 1 sec & $<$ 1 sec\\\hline
    %brock800\_1.clq & 800 & 207505 & 790 & 1 min 54 sec & 32 sec\\\hline
    %brock800\_2.clq & 800 & 208166 & 790 & 1 min 54 sec & 30 sec\\\hline
    %brock800\_3.clq & 800 & 207333 & 789 & 1 min 50 sec & 31 sec\\\hline
    %brock800\_4.clq & 800 & 207643 & 790 & 1 min 51 sec & 30 sec\\\hline
    %p\_hat300-1.clq & 300 & 10933 & 261 & 1 min 25 sec & 41 sec\\\hline
    %p\_hat300-2.clq & 300 & 21928 & 273 & 2 sec & $<$ 1 sec\\\hline
    %p\_hat300-3.clq & 300 & 33390 & 291 & $<$ 1 sec & $<$ 1 sec\\\hline
    p\_hat500-1.clq & 500 & 31569 & 450 & 3 hr 48 min & 1 hr 23 min \\\hline
    %p\_hat500-2.clq & 500 & 62946 & 464 & 40 sec & 12 sec\\\hline
    %p\_hat500-3.clq & 500 & 93800 & 490 & $<$ 1 sec & $<$ 1 sec\\\hline
    p\_hat700-1.clq & 700 & 60999 & 635 & $>$ 1 week & 93 hr 20 min \\\hline
    p\_hat700-2.clq & 700 & 121728 & 651 & 15 min 10 sec & 3 min 44 sec\\\hline
    %p\_hat700-3.clq & 700 & 183010 & 690 & 20 sec & 6 sec\\\hline
    %p\_hat1000-1.clq & 1000 & 122253 & ? & ? & ?\\\hline
    p\_hat1000-2.clq & 1000 & 244799 & 946 & 31 hr 26 min & 5 hr 28 min\\\hline
    %p\_hat1000-3.clq & 1000 & 371746 & 989 & 2 min 47 sec & 48 sec\\\hline
    %p\_hat1500-1.clq & 1500 & 284923 & ? & ? & ?\\\hline
    %p\_hat1500-2.clq & 1500 & 568960 & ? & ? & ?\\\hline
    p\_hat1500-3.clq & 1500 & 847244 & 1488 & 20 min 57 sec & 5 min 3 sec\\\hline
\end{tabular}
\end{center}\vspace{-0.5cm}
\end{table}

\HYBRIDVCPARM\ is a parameterized hybrid version that does not take advantage
of vertex folding, while \HYBRIDVCFPARM\ is a parameterized version implemented
using the coloring technique described in the previous section for folding.
In all the conducted experiments, the folding technique
is at most two times slower than the simple generic branching algorithm. It gets faster
as the difference between the highest and lowest vertex-degrees gets smaller. In
particular, applying vertex folding via our coloring technique, is much faster on
regular graphs. To illustrate,
we report experiments on a well known 4-regular graph (the 120-Cell on 300 vertices),
by varying the input parameter, and results are reported in Table \ref{tab:vctable2}.

\begin{table}[htb]
\caption{\label{tab:vctable2}\HYBRIDVCPARM\ vs. \HYBRIDVCFPARM\ (with folding) on a 4-regular graph having 300 vertices and 600 edges.}
\begin{center}
\begin{tabular}{| c | c | c | c |} \hline
    Vertex Cover Size ($k$) ~~&~ Answer ~&~~ No Folding ~~&~~ With Folding ~~\\\hline
    %192 & yes & 14 sec & $<$ 1 sec\\\hline
    %191 & yes & 17 sec & 6 sec\\\hline
    190 & yes & 2 hr 14 min & 6 min 27 sec\\\hline
    165 & no & $>$ 4 days & 46 min 56 sec\\\hline
    160 & no & 38 hr 2 min & 2 min 32 sec\\\hline
\end{tabular}
\end{center}\vspace{-0.5cm}
\end{table}

As for the \DSFULL\ problem, \ALDSOPT\ denotes the optimization version using the
adjacency-lists representation and \HYBRIDDSOPT\ the optimization version using the
hybrid graph representation. Running times on random graphs,
with various densities,
are given in Table \ref{tab:dstable1}.

\begin{table}[htb]
\caption{\label{tab:dstable1}\ALDSOPT\ vs. \HYBRIDDSOPT.}
\begin{center}
\begin{tabular}{| c | c | c | c | c | c |} \hline
    Graph ~&~ $|V|$ ~&~ $|E|$ ~&~ $|D|$ ~&~ \ALDSOPT\ &\HYBRIDDSOPT\ ~\\\hline
    rgraph1 & 100 & 400 & 16 & 21 min 4 sec & 4 min 11 sec\\\hline
    %rgraph2 & 100 & 600 & 11 & 5 min 5 sec & 53 sec\\\hline
    %rgraph3 & 100 & 1500 & 6 & 41 sec & 5 sec\\\hline

    rgraph4 & 150 & 1200 & 14 & 16 hr 46 min & 2 hr 27 min\\\hline
    rgraph5 & 150 & 1500 & 11 & 3 hr 31 min & 28 min 20 sec\\\hline
    %rgraph6 & 150 & 3000 & 6 & 2 min 8 sec & 12 sec\\\hline

    rgraph7 & 150 & 3000 & 7 & 27 min 16 sec & 2 min 1 sec\\\hline

    rgraph8 & 200 & 4500 & 9 & 5 hr 44 min & 30 min 8 sec\\\hline
    rgraph9 & 200 & 5000 & 8 & 1 hr 20 min & 6 min 46 sec\\\hline
    rgraph10 & 200 & 6000 & 6 & 1 hr 36 min & 7 min 13 sec\\\hline
    %rgraph11 & 200 & 12000 & 4 & 6 min 49 sec & 17 sec\\\hline

    rgraph12 & 250 & 9000 & 8 & 14 hr 37 min & 56 min 53 sec\\\hline
    rgraph13 & 250 & 10000 & 7 & 1 hr 30 min & 5 min 34 sec\\\hline
    rgraph14 & 250 & 12000 & 5 & 4 hr 41 min & 16 min 19 sec\\\hline
    %rgraph15 & 250 & 24000 & 3 & 19 sec & $<$ 1 sec\\\hline

    %rgraph16 & 300 & 22461 & 4 & 8 min 31 sec & 17 sec\\\hline
    rgraph17 & 300 & 22258 & 4 & 28 min 32 sec & 1 min 10 sec\\\hline
    rgraph18 & 300 & 11063 & 8 & 133 hr 38 min & 5 hr 54 min \\\hline
    rgraph19 & 300 & 11287 & 8 & $>$ 7 days & 8 hr 14 min\\\hline
    %rgraph20 & 1000 & 374633 & 3 & 4 min 37 sec & 2 min 29 sec\\\hline
    rgraph21 & 1000 & 374552 & 3 & 28 min 37 sec & 6 min 36 sec\\\hline
\end{tabular}
\end{center}\vspace{-0.5cm}
\end{table}

To demonstrate the efficiency of the edge addition operation, we implemented two
versions of the parameterized \CEFULL\ algorithm described in Section 2.
\ALCEPARM\ denotes the implementation using the
adjacency-lists representation and \HYBRIDCEPARM\ refers to the version
implemented with the hybrid graph
representation. The running times of the implementations on random graphs
are shown in Table \ref{tab:cetable1}. The graphs were generated based on
fixing the number of vertices, the number of clusters ($C$ in the table below)
and the parameter $k$. For each such triple, the vertices were distributed
randomly over the $C$ clusters, then intra-cluster and inter-cluster
edges were deleted and added (respectively)
randomly to obtain a \CEFULL\ instance.

\begin{table}[htb]
\caption{\label{tab:cetable1}\ALCEPARM\ vs. \HYBRIDCEPARM.}
\begin{center}
\begin{tabular}{| c | c | c | c | c | c | c |}  \hline
    Graph ~&~ $|V|$ ~&~ $|E|$ ~&~ C ~&~ k ~&~ \ALCEPARM\ &\HYBRIDCEPARM\ ~\\\hline
    %rgraph1 & 125 & 1508 & 5 & 10 & 0.42 sec & 0.15 sec\\\hline
    %rgraph1 & 125 & 1508 & 5 & 15 & 15.18 sec & 4.31 sec\\\hline
    rgraph1 & 125 & 1508 & 5 & 20 & 7 min 6 sec & 2 min 12 sec\\\hline
    rgraph1 & 125 & 1508 & 5 & 25 & 4 hr 15 min & 1 hr 12 min\\\hline
    rgraph1 & 125 & 1508 & 5 & 30 & 5 days 21 hr & 1 day 2 hr\\\hline

    %rgraph2 & 250 & 3010 & 10 & 10 & 0.50 sec & 0.20 sec\\\hline
    %rgraph2 & 250 & 3010 & 10 & 15 & 16.74 sec & 4.80 sec\\\hline
    rgraph2 & 250 & 3010 & 10 & 20 & 8 min 54 sec & 2 min 22 sec\\\hline
    rgraph2 & 250 & 3010 & 10 & 25 & 4 hr 46 min & 1 hr 15 min\\\hline
    rgraph2 & 250 & 3010 & 10 & 30 & 6 day 9 hr & 1 day 10 hr\\\hline

    %rgraph3 & 500 & 6013 & 20 & 10 & 1.55 sec & 0.58 sec\\\hline
    %rgraph3 & 500 & 6013 & 20 & 15 & 52.10 sec & 16.18 sec\\\hline
    rgraph3 & 500 & 6013 & 20 & 20 & 28 min 28 sec & 8 min 46 sec\\\hline
    rgraph3 & 500 & 6013 & 20 & 25 & 15 hr 13 min & 4 hr 25 min\\\hline
    rgraph3 & 500 & 6013 & 20 & 30 & 21 days 6 hr & 4 days 2 hr\\\hline

    %rgraph4 & 250 & 6139 & 5 & 10 & 1.4 sec & 0.27 sec\\\hline
    %rgraph4 & 250 & 6139 & 5 & 15 & 52.51 sec & 8.90 sec\\\hline
    rgraph4 & 250 & 6139 & 5 & 20 & 28 min 31 sec & 4 min 36 sec\\\hline
    rgraph4 & 250 & 6139 & 5 & 25 & 15 hr 13 min & 2 hr 26 min\\\hline
    rgraph4 & 250 & 6139 & 5 & 30 & 20 days 14 hr & 3 days 6 hr\\\hline
\end{tabular}
\end{center}\vspace{-0.5cm}
\end{table}

All codes were implemented in standard C, and experiments were run
on three types of conventional machines:
Intel Core2 Duo 2.33 GHz, Intel Xeon Processor X5550 2.66 GHz Quad
Core, and Intel Core i7-2600 3.40GHz Quad Core. However, the numbers
reported in each row were obtained on the same architecture.

\section{Conclusion}\label{sec-conclusion}
We presented a hybrid graph representation that efficiently trades space for time
and facilitates many common graph operations required during recursive backtracking.
Experiments on \VCFULL, \DSFULL\ and \CEFULL\ showed the utility of using this
dynamic data structure. The running times of the same algorithm were shown to be
consistently reduced, sometimes from days to hours.

The hybrid method can also be applied to recursive enumeration problems that
are based on search-tree algorithms, such as the well known
Bron-Kerbosch algorithm
for \MCEFULL~\cite{BK73}. Some of the
presented implementation techniques are used in our recent implementations
of the algorithms described in~\cite{RTST}.

The main focus in this paper was on graph modification operations that reduce
the original graph size, such as vertex deletion and edge contraction.
This can be applied for the implementation of branch-and-reduce algorithms.
Operations that increase the size of a graph are harder to implement
using the presented techniques.
Vertex addition and non-permanent edge addition
are notable examples that remain to be considered.

%% --------------------------------------------------------------------
%       Bibliography
%% --------------------------------------------------------------------

\bibliography{references}
\bibliographystyle{abbrv}
\end{document}